# Operational regimes of lasers based on gain media with a large Raman scattering cross-section

E. A. Tereschenkov[1,2,3], E. S. Andrianov[1,2,3], A. A. Zyablovsky[1,2,3,4], A. A. Pukhov,[1,3] A. P. Vinogradov[1,2,3] & A. A. Lisyansky[5,6]

[1]Moscow Institute of Physics and Technology, 141700, 9 Institutskiy per., Moscow, Russia

[2]Dukhov Research Institute of Automatics (VNIIA), 127055, 22 Sushchevskaya, Moscow, Russia

[3]Institute for Theoretical and Applied Electromagnetics, 125412, 13 Izhorskaya, Moscow, Russia

[4]Kotelnikov Institute of Radioengineering and Electronics RAS, 11-7 Mokhovaya, Moscow 125009, Russia

[5]Department of Physics, Queens College of the City University of New York, Flushing, New York 11367, U.S.A.

[6]The Graduate Center of the City University of New York, New York, New York 10016, U.S.A.

**We report on unusual regimes of operation of a laser with a gain medium with a large Raman scattering cross-section, which is often inherent in new types of gain media such as colloidal and epitaxial quantum dots and perovskite materials. These media are characterized by a strong electron-phonon coupling. Using the Fröhlich Hamiltonian to describe the electron-phonon coupling in such media, we analyze the operation of the system above the lasing threshold. We show that below a critical value of the Fröhlich constant, the laser can only operate in the conventional regime: namely, there are coherent cavity photons but no coherent phonons. Above the critical value, a new pump rate threshold appears. Above this threshold, either joint self-oscillations of coherent phonons in the gain medium and photons in a cavity or a chaotic regime are established. We also find a range of the values of the Fröhlich constant, the pump rate, and the resonator eigenfrequency, in which more than one dynamical regime of the system is stable. In this case the laser dynamics is determined by the initial values of the resonator field, the active medium polarization, the population inversion, and phonon amplitude.**

**Introduction**

Recent advances in chemistry and nanotechnology have significantly expanded the types of active laser media. At present, active media based on colloidal[1,2] and epitaxial quantum dots (QDs)[3-5], perovskites[6-8], and dye molecules[9-11] are characterized by a high gain and are widely used for the miniaturization of lasers[12-16]. As a rule, the interaction between electrons and optical phonons is strong in active media with large Raman cross-section[17-19]. In this case, it becomes possible to amplify phonons through the action of the electromagnetic field on the electronic subsystem. Thus, in addition to high light amplification, systems with a strong electron-phonon coupling can effectively amplify optical phonons with a short wavelength and create phonon nanolasers[20]. This is facilitated by the fact that the wavelength of an optical phonon at terahertz frequencies is equal to several nanometers, which corresponds to the typical length scale of nanostructures[21]. Since phonons have a relatively high mean free path (of 100–1000 μm)[22] in the terahertz range, they can provide high resolution when used for image processing and nanostructure research.

In this work, we focus on the influence of nuclear vibrations in materials with a strong electron-phonon coupling upon the amplifying properties of conventional lasers. To be specific, we consider amplifying media based on QDs. In such lasers, the pumping is assumed to be incoherent, and lasing is determined by the interaction of the polarization of the QD electronic subsystem with the resonator mode. The coupling of the cavity electric field and the active medium polarization is treated in the dipole approximation and is characterized by the Rabi constant[23,24]. As a rule, the observed gain in conventional lasers is much greater than the gain due to the stimulated Raman scattering (SRS). Therefore, even in lasers using active media with a strong interaction between electrons and phonons, the influence of optical phonons on laser operation is usually neglected.

We describe the interaction between the electrons and optical phonons via the Fröhlich Hamiltonian[18,25-28]. Because of the strong coupling of optical phonons with the electric field, the Fröhlich constant of the interaction, $g$, may be comparable or even larger than the Rabi constant.

We use computer simulation to show that there is a region bounded by the curve $g = g_{cr}(D_0)$ in the plane of parameters $(g, D_0)$, where $D_0$ is the population inversion characterizing the incoherent pump. Inside this region, only usual lasing regimes are possible - these are the trivial regime without the coherent optical field and the lasing regime, in which the mean coherent optical field is not zero. In both regimes, coherent phonons are not excited.

We focus our attention on the regions in which $D_0$ is greater than the laser threshold $D_{th\_opt}$ and the Rabi frequency is much less than optical frequencies. Outside of the region $g < g_{cr}(D_0)$, new operational regimes appear. One of these new regimes is characterized by the excitation of *coherent* joint self-oscillations of both an optical field in the cavity and optical phonons in the gain medium.

We also find that for QDs with a low transverse relaxation rate, a regime with coherent phonons may be transformed into a chaotic regime with a spectrum containing an infinite number of frequencies incommensurable with optical phonon frequencies.

Finally, we show that in contrast to the conventional lasers, there exists a region of values of $g$ and $D_0$ values in which two different regimes are stable simultaneously. These regimes may be any pair of the following: the trivial non-lasing regime, the conventional laser regime, the lasing regime accompanied by coherent phonons, and the chaotic regime. Regime which is realized in practice is determined by the initial conditions.

**Model**

We study the dynamics of a laser consisting of a single-mode cavity with the frequency $\omega_a$ and QDs serving as an active medium. In the QD, the electronic degree of freedom is usually an exciton excitation. The exciton interacts with the quantized vibrations of the QD nuclei – optical phonons; we consider one optical phonon mode with the eigenfrequency $\omega_b$. Since underlying lasing is a resonant phenomenon, we can neglect the excitation of off-resonant levels and only consider two electronic states of the active medium[29] – the ground state $|g\rangle$, which is in the valence band, and the excited state $|e\rangle$, which is in the conduction band. The transition frequency $\omega_\sigma$ between these states corresponds to the excitation of an exciton.

In colloidal QDs and dyes in the optical range, the transverse relaxation rate $\gamma_\sigma$ is about $0.001 - 0.01 eV$ [30-32]. This is usually much greater than the decay rate in the resonator $\gamma_a$. The relationship between the relaxation rates $\gamma_a \ll \gamma_\sigma$ takes place in gas lasers[29,33]. We, therefore, can use the gas-laser model to describe the QD-based lasers.

We emphasize that we consider a conventional laser scheme that differs from the Raman laser. As opposed to the case considered here, in the latter, pumping is coherent, and the laser must have at least two cavity modes with the frequency difference $\Delta\omega$ equal to the phonon frequency $\omega_b$. Because the nonlinear Fröhlich interaction provides the energy flow from the mode with higher frequency to the mode with lower frequency, one does not have to create population inversion in the active medium.

Below we consider a laser whose cavity size $L$ is small enough to ensure inequality $c/L \gg \omega_b$. Since $\Delta\omega \sim c/L$, this inequality provides $\Delta\omega \gg \omega_b$, and therefore, the effect of Raman laser can be neglected. In this case, only one cavity mode is involved in the laser dynamics, and population inversion is necessary. Below we assume that $L \ll c/\omega_b \sim 100\mu\text{m}$.

Before proceeding to the study of the effect of the electron-phonon interaction on the operation of a laser, we recall the main points of the theory of lasers[24,34] and introduce the corresponding notations.

The system Hamiltonian is the usual Jaynes-Cummings Hamiltonian with two additional terms describing the energy of intermolecular vibrations of nuclei and the Fröhlich Hamiltonian of the interaction of the electronic and vibrational subsystems[18,25-28]:

$$\hat{H} = \hbar\omega_a \hat{a}^\dagger \hat{a} + \hbar\omega_\sigma \hat{\sigma}^\dagger \hat{\sigma} + \hbar\Omega_R\left(\hat{a}\hat{\sigma}^\dagger + \hat{a}^\dagger \hat{\sigma}\right) + \hbar\omega_b \hat{b}^\dagger \hat{b} + \hbar g \hat{\sigma}^\dagger \hat{\sigma}\left(\hat{b}^\dagger + \hat{b}\right). \tag{1}$$

In Eq. (1), the first term is the Hamiltonian of the electromagnetic field in the cavity, the operators $\hat{a}^\dagger$ and $\hat{a}$ are the operators of creation and annihilation of a quantum of the electromagnetic field. These operators satisfy the commutation relation $[\hat{a}, \hat{a}^\dagger] = \hat{1}$. The second term describes the Hamiltonian of the QD exciton modeled by a two-level system. The operators $\hat{\sigma} = |g\rangle\langle e|$ and $\hat{\sigma}^\dagger = |e\rangle\langle g|$ are transitions operators from the excited state to the ground state and vice versa, and $\hat{\sigma}^\dagger \hat{\sigma}$ is the operator of the population of the excited electronic state in the QD. These operators satisfy the commutation relation $[\hat{\sigma}^\dagger, \hat{\sigma}] = \hat{D}$, where the operator $\hat{D}$ determines the difference between the populations of the excited and ground states. The third term is the interaction of the QD with the cavity mode[24], $\Omega_R = -\mathbf{d}_{eg} \cdot \mathbf{E}(\mathbf{r})/\hbar$ is the Rabi constant of the interaction, $\mathbf{d}_{eg}$ is the matrix element of exciton transition, and $\mathbf{E}(\mathbf{r})$ is the electric field "per one photon" of the cavity mode at the QD location $\mathbf{r}$. The fourth term in Hamiltonian (1) describes the optical phonon in the harmonic approximation, $\hat{b}^\dagger$ and $\hat{b}$ are the operators of creation and annihilation of the phonon; these operators satisfy the commutation relation $\left[\hat{b},\hat{b}^\dagger\right]=\hat{1}$. The last term is the interaction of the electronic and vibrational subsystems[18,25-28]. The operator $\hat{b}^\dagger+\hat{b}$ has the meaning of the operator of the amplitude of the nuclear excitation. The Fröhlich constant can be evaluated employing the expression for the Raman scattering cross-section at the Stokes frequency[18]:

$$\sigma_{Cross\_section} = \frac{4\pi}{3} \frac{|\mathbf{d}_{eg}|^4 \omega_{St}^4 g^2}{c^4 \hbar^2 (\omega-\omega_\sigma+i\gamma_\sigma)^2 (\omega_\sigma-\omega_{St})^2}. \qquad (2)$$

The experimentally measured value of $\sigma_{Cross\_section}$ for CdSe QDs is $\sigma_{Cross\_section} \approx 4\cdot 10^{-6}\,\text{Å}^2$ (see Ref. 19). Using the values of the phonon frequency $\omega_b \approx 0.025\,\text{eV}$ (which corresponds to the optical phonon in CdSe[35]) and the QD dipole moment $d_{eg} \approx 5-10\,\text{D}$, we obtain that the Fröhlich constant is of the order of $g \simeq 10^{-2}\,\text{eV}$ which we use in our study.

Thus, Hamiltonian (1) is applicable for systems consisting of active medium with high Raman scattering cross-section placed in the cavity. An example of such an active medium is CdSe, CdS, and PbS quantum dots. Note that, for simplicity, we consider only one optical phonon mode in an active medium.

To describe relaxation processes, it is necessary to introduce reservoirs[36,37] which interact with electronic and phonon subsystems of the QD and the cavity mode. We consider reservoirs describing Ohmic losses in the cavity, energy relaxation of the exciton, energy relaxation of optical phonons, and the phonon reservoir responsible for exciton dephasing. Eliminating the reservoir degrees of freedom in the Born-Markov approximation results in the master equation in the Lindblad form[36,37] (for details, see **Methods. Master equation for the system density matrix**).

Employing Lindblad equation, one can obtain the equation of motion for the expected value of the involved operators. To find the operation regimes, we study the evolutions of the expected values of the following operators: the annihilation operator of the field in the resonator

$\langle \hat{a} \rangle = a$, which is proportional to the average field in the resonator, the operator of the exciton total dipole moment, $\sigma = \langle \hat{\sigma} \rangle$, the operator of the population inversion $\langle \hat{D} \rangle = D$ of the excited level, and the annihilation operator of optical phonon $b = \langle \hat{b} \rangle$. As a result, we arrive at the following equations of motion, which are the expanded Maxwell-Bloch equations (see **Methods. Derivation of Eqs. (3) – (6)**):

$$da/dt = (-i\omega_a - \gamma_a/2)a - i\Omega_R \sigma, \tag{3}$$

$$d\sigma/dt = \left(-i\left(\omega_\sigma + g(b+b^*)\right) - \gamma_\sigma/2\right)\sigma + i\Omega_R aD, \tag{4}$$

$$db/dt = (-i\omega_b - \gamma_b/2)b - ig(D+1)/2, \tag{5}$$

$$dD/dt = -(\gamma_p + \gamma_D)(D - D_0) + 2i\Omega_R(a^\dagger \sigma - \sigma^\dagger a). \tag{6}$$

The quantities $\gamma_a$, $\gamma_D$, and $\gamma_p$ have the meaning of the rates of respective relaxation and pump processes, $\gamma_\sigma = \gamma_D + \gamma_p + 2\gamma_{deph}$ where $\gamma_{deph}$ is the dephasing rate[37]. In Eq. (6), $D_0 = (\gamma_p - \gamma_D)/(\gamma_p + \gamma_D)$. We have neglected the correlations between the operators $\hat{a}$ and $\hat{\sigma}$ and the operators $\hat{a}$ and $\hat{D}$, assuming that $\langle \hat{a}\hat{D} \rangle \simeq \langle \hat{a} \rangle \langle \hat{D} \rangle$ and $\langle \hat{a}^\dagger \hat{\sigma} \rangle \simeq \langle \hat{a}^\dagger \rangle \langle \hat{\sigma} \rangle$. In addition, we have neglected the correlation between operators $\hat{b}$ and $\hat{\sigma}$ assuming that $\langle (\hat{b}+\hat{b}^\dagger)\hat{\sigma} \rangle \simeq \langle \hat{b}+\hat{b}^\dagger \rangle \langle \hat{\sigma} \rangle$. By splitting the correlators, we neglect quantum fluctuations; in particular, we do not explicitly take into account spontaneous emission. This is correct if we consider the laser dynamics well below or well above the threshold [18,23,34,36].

Solving Eqs. (3) – (6) does not allow for a detailed description of the spontaneous emission process[38] but describes rather well the induced emission[23,37]. Note that Eqs. (3) – (6) describe an indirect coupling between the cavity electric field of the cavity mode and the optical phonon of the active medium. Indeed, Eq. (3) includes the interaction of $a$ with $\sigma$ through the term $-i\Omega_R \sigma$, while Eq. (4) has the term $-ig(b+b^*)\sigma$ that couples $\sigma$ with $b$. The direct interaction is absent because we assume that QDs are Raman active.

Below, we describe different operational regimes of a laser by three parameters of the problem. The first parameter is the Fröhlich interaction constant $g$. At $g = 0$, we deal with a conventional laser. The second relevant parameter is the pump rate $\gamma_p$. This parameter determines the transition to the lasing regime. As commonly accepted, we describe the pump rate with the parameter $D_0 = (\gamma_p - \gamma_D)/(\gamma_p + \gamma_D)$, which is uniquely determined by $\gamma_p$. The third parameter of our problem is resonator eigenfrequency $\omega_a$.

**Results**

**The regime of convention laser without excitation of coherent phonons**

The trivial solution that corresponds to the absence of coherent oscillations of $a, \sigma, D$, and $b$ has the form

$$a_{triv} = 0, \tag{7}$$

$$\sigma_{triv} = 0, \tag{8}$$

$$D_{triv} = D_0, \tag{9}$$

$$b_{triv} = \frac{-g(D_0 + 1)}{2(\omega_b - i\gamma_b/2)}. \tag{10}$$

This solution is analogous to the trivial solution for a laser, in which the phonon degree of freedom of the active medium molecule is not taken into account. Usually, this solution is stable below the lasing threshold. In the case under consideration, the trivial solution is characterized by a constant expected value of the phonon operator $\hat{b}$, which corresponds to a constant displacement of the nuclei.

By increasing the pumping $D_0$, we expect, besides the trivial solution, to find a self-oscillating solution. Our computer simulation shows no coherent phonons in this regime, and the coherent part of $b$, which we denote as $b_{ph}$, is still equal to zero.

Since we are interested in the stationary behavior of the system, namely, the behavior at large times, $t \gg \gamma_a^{-1}, \gamma_\sigma^{-1}, \gamma_b^{-1}$, we search for self-oscillating solutions in the trial form:

$$a = a_{opt}(D_0)\exp(-i\omega_{gen}t), \tag{11}$$

$$\sigma = \sigma_{opt}(D_0)\exp(-i\omega_{gen}t), \tag{12}$$

$$D = D_{th\_opt}, \tag{13}$$

$$b = \frac{-g(D_{th\_opt} + 1)}{2(\omega_b - i\gamma_b/2)} \tag{14}$$

with still unknown frequency $\omega_{gen}$ and the laser threshold $D_{th\_opt}$. Here $a_{opt}$ and $\sigma_{opt}$ are time-independent amplitudes of oscillations of $a$ and $\sigma$. Thus, according to Eqs. (9) and (13)

$$D = D_0 \text{ if } D_0 < D_{th\_opt} \text{ and } D = D_{th\_opt} \text{ if } D_0 > D_{th\_opt}. \tag{15}$$

Substituting Eqs. (11)-(13) into Eqs. (3)-(6) and equating the real and imaginary parts of Eq. (4) to zero, we obtain the following expression for the population inversion $D_{th\_opt}$ and the self-oscillation frequency $\omega_{gen}$:

$$D_{th\_opt} = \frac{\left(-\delta_{gen}\Delta_{gen} + \frac{\gamma_\sigma \gamma_a}{4}\right)\left(\omega_b^2 + \frac{\gamma_b^2}{4}\right) - \omega_b g^2 \Delta_{gen}}{\omega_b g^2 \Delta_{gen} + \Omega_R^2\left(\omega_b^2 + \frac{\gamma_b^2}{4}\right)}, \tag{16}$$

$$\left(-\delta_{gen}\frac{\gamma_a}{2} - \Delta_{gen}\frac{\gamma_\sigma}{2}\right)\left(\omega_b g^2 \Delta_{gen} + \Omega_R^2\left(\omega_b^2 + \frac{\gamma_b^2}{4}\right)\right) - \omega_b g^2 \frac{\gamma_a}{2}\left(-\delta_{gen}\Delta_{gen} + \frac{\gamma_\sigma \gamma_a}{4} + \Omega_R^2\right) = 0, \tag{17}$$

where $\Delta_{gen} = \omega_{gen} - \omega_a$ and $\delta_{gen} = \omega_{gen} - \omega_\sigma$. Equations (16) and (17) have terms proportional to $g^2$, resulting in the dependence of the stationary population inversion and the frequency of self-oscillations on the Fröhlich constant $g$.

By introducing the notation

$$\omega_\sigma^{(g)}(D_{th\_opt}) = \omega_\sigma - \frac{\omega_b g^2 (D_{th\_opt} + 1)}{\omega_b^2 + \gamma_b^2/4} \tag{18}$$

we can reduce Eq. (17) to the form

$$\omega_{gen} = \frac{\omega_\sigma^{(g)}(D_{th\_opt})\gamma_\sigma + \omega_a \gamma_a}{\gamma_\sigma + \gamma_a}. \tag{19}$$

Note that $\omega_{gen}$ defined by Eq. (19) is similar to the frequency pulling in the absence of the interaction between optical phonon and exciton (see Eq. (44) in **Methods. Maxwell-Bloch equations in the absence of interaction between exciton and optical phonon**).

Comparing Eq. (19) with Eq. (51) in **Methods. Effective change in the exciton frequency due to a constant displacement of the QD nuclei under incoherent pumping** shows that $\omega_\sigma^{(g)}$ is the effective transition frequency of an exciton under the influence of incoherent pumping due to the interaction with an optical phonon.

For the field amplitude in the resonator and the polarization, we obtain the following expressions:

$$a_{opt} = \sqrt{\frac{(\gamma_p + \gamma_D)(D_0 - D_{th\_opt})}{2\gamma_a}},$$
$$\sigma_{opt} = \frac{i\Delta - \gamma_a/2}{i\Omega_R}\sqrt{\frac{(\gamma_p + \gamma_D)}{2\gamma_a}(D_0 - D_{th\_opt})}. \tag{20}$$

The quantity $D_{th\_opt}$ has the meaning of the generation threshold for self-oscillation of the cavity field. Such a behavior corresponds to the Hopf bifurcation[39].

**The laser regime with excitation of coherent phonons**
In this section, we show that the interaction of an exciton with the optical phonon may lead to the appearance of either a regime with coherent phonons or deterministic chaos.

**The resonant case, $\omega_a = \omega_\sigma^{(g)}$**

Let us first consider the resonant case. For this purpose we choose the value of $\omega_a$ to be equal to the effective exciton frequency $\omega_\sigma^{(g)}$, defined in Eq. (18), for any value of $g$ and $D_0 > D_{th\_opt}$. In this case, we have two parameters $g$ and $D_0$ that determine operational regimes.

The possible dynamical regimes found with the aid of computer simulations of Eqs. (3) – (6) are shown on plane $(g, D_0)$, in Fig. 1. It turns out that there are two threshold values of pumping $D_0$, which separate different regimes. When the pumping $D_0$ is less than the optical threshold $D_{th\_opt}$ determined by Eq. (16), computer simulation shows that only trivial solution (7) – (10) exists and is stable (the blue area in Fig. 1). Note that at large pumping, disruption of generation occurs. This is due to the increase in the dissipation rate of the exciton polarization $\gamma_\sigma$ with an increase in the pump rate $\gamma_p$. When the pump rate becomes as high as $\gamma_p \gg \gamma_D, \gamma_{deph}$, the dissipation of the polarization becomes so large that laser generation is hindered[33,40].

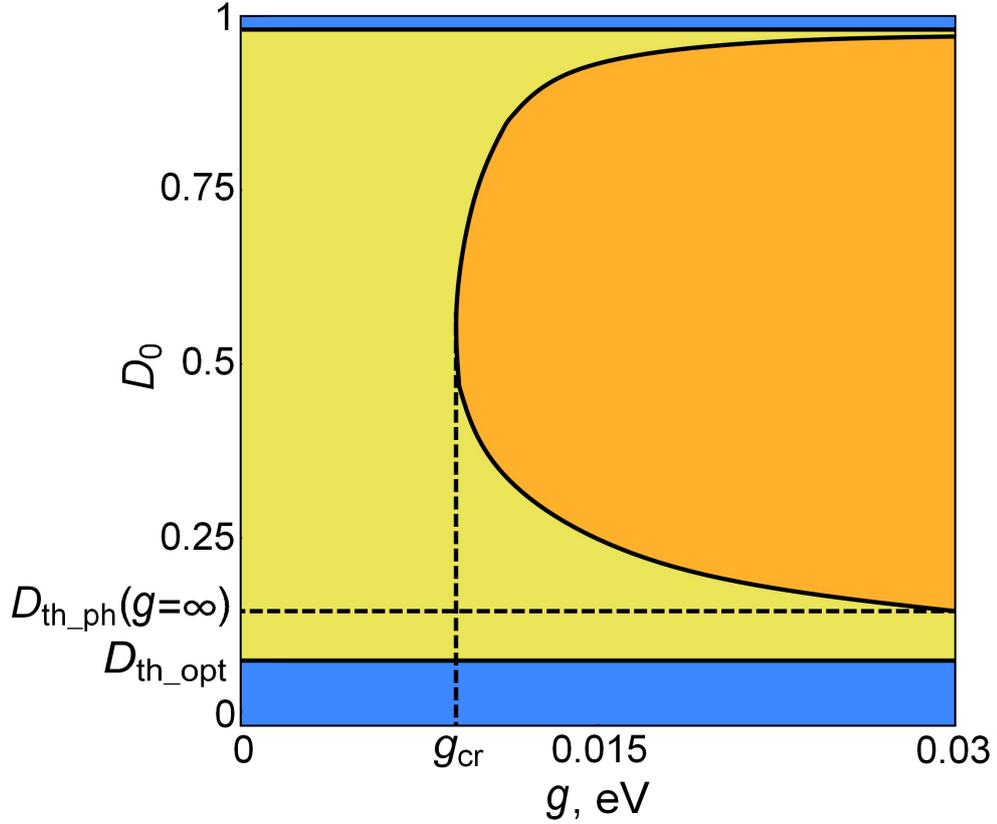

**Figure 1**. Operational regimes in the plane of parameters $(g, D_0)$. There is no lasing in the blue region ($D_0 < D_{th\_opt}(g)$), in which only the trivial solution is stable. In the upper blue area, the generation is disrupted by large pumping. The yellow region is the self-oscillating laser regime without excitation of coherent phonons. The orange region shows the regime of self-oscillations of coherent phonons and the periodic-in-time part of the optical field in the cavity. The system parameters are $\Omega_R = 0.01 eV$, $\gamma_b = 0.0002 eV$, $\gamma_a = 0.001 eV$, $\omega_b = 0.01 eV$, $\omega_\sigma = 2.4 eV$, $\gamma_{deph} = 0.005 eV$, and $\gamma_D = 0.001 eV$.

When $D_0 > D_{th\_opt}$, numerical analysis shows of Eqs. (3) – (6) that the trivial solution becomes unstable, and a new self-oscillating solution without coherent phonons arises (see Eq. (20)). This solution is stable for any $g$ until $D_0 < D_{th\_ph}(g = \infty)$ (the yellow region in Fig. 1). In this regime, the expected value of the nuclear displacement due to their coherent vibrations about equilibrium positions is zero, $b_{ph} = 0$. Therefore, despite accounting for the electron-phonon interaction, the laser radiation spectrum contains neither Stokes nor anti-Stokes frequencies. It means that nuclei do not take part in self-oscillation. This is exactly what is usually assumed when neglecting the electron-phonon interaction.

In the region $D_0 > D_{th\_ph}(g = \infty)$, our computer simulation of Eqs. (3) – (6) shows that there is a curve $D_0 = D_{th\_ph}(g)$ of the Hopf bifurcations (the solid curve in Fig. 1). This curve bounds the area (colored in orange in Fig. 1) in which the lasing regime without coherent phonons becomes unstable, and as we show below, a new regime with coherent phonons arises. We also show that this is a new Hopf bifurcation leading to the formation of new self-oscillation at the frequency close to $\omega_v$.

To study the latter regime, it is convenient to present the trial form for the expected values of the operators $\hat{a}$, $\hat{\sigma}$, $\hat{D}$, and $\hat{b}$ in the following form:

$$a(t) = (a_{opt} + a_{ph}(t))\exp(-i\omega_{gen}t), \tag{21}$$

$$\sigma(t) = (\sigma_{opt} + \sigma_{ph}(t))\exp(-i\omega_{gen}t), \tag{22}$$

$$D(t) = D_{th\_opt} + D_{ph}(t), \tag{23}$$

$$b(t) = \frac{-g(D_{th\_opt} + 1)}{2(\omega_b - i\gamma_b/2)} + b_{ph}(t). \tag{24}$$

Here, we split the amplitudes of $a(t)$, $\sigma(t)$, $D(t)$ and $b(t)$ into the time-independent parts ($a_{opt}$, $\sigma_{opt}$, $D_{th\_opt}$, and $\frac{-g(D_{th\_opt}+1)}{2(\omega_b - i\gamma_b/2)}$), corresponding to the conventional laser regime (see Eqs. (11) - (14)), and the time-dependent parts ($a_{ph}(t)$, $\sigma_{ph}(t)$, $D_{ph}(t)$ and $b_{ph}(t)$), forming new non-harmonic self-oscillations with the period $T = 2\pi/\omega_{b,gen}$. The lowest frequency $\omega_{b,gen}$ is the frequency of the excited coherent phonons $\omega_{b,gen} \equiv 2\pi/T$. Note that $\omega_{b,gen} \neq \omega_b$.

We solve Eqs. (3) - (6) numerically for different values of $g$ and $D_0$ and consider the presence of time-dependent parts mentioned above.

In order to characterize the transition to the self-oscillation with coherent phonons, we average the quantities $|a_{ph}(t)|$, $|\sigma_{ph}(t)|$, $|D_{th}(t)|$, and $|b_{th}(t)|$ over the period of the self-oscillations: $\langle...\rangle = \frac{1}{T}\int_{t_0}^{t_0+T}|...|dt$, where, $t_0 \gg \gamma_a^{-1}, \gamma_\sigma^{-1}, \gamma_b^{-1}$ is the time during which the system

reaches stationary oscillations, $T$ is the period of new self-oscillations. Figure 2 shows the dependence of $\langle |b_{ph}| \rangle$ on the pump rate $D_0$ for fixed $g > g_{cr}$. This dependence behaves like $\sim \sqrt{D_0 - D_{th\_ph}}$, which shows that the second Hopf bifurcation occurs at the second threshold $D_{th\_ph}$.

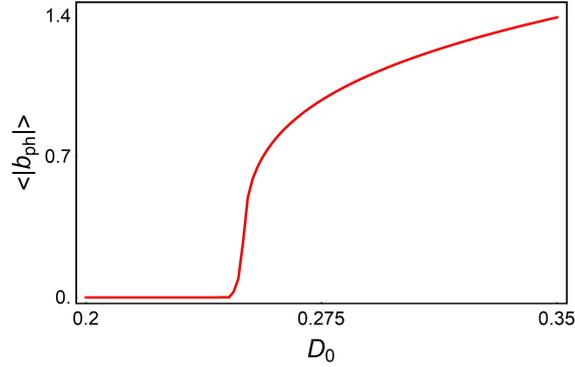

**Figure 2**. The dependence of $\langle |b_{ph}| \rangle$ on the pump rate $D_0$ for fixed $g = 0.015\text{eV} > g_{cr}$. Other parameters are the same as in Fig.1.

In the new regime, the spectrum of the self-oscillations is not characterized by a single frequency $\omega_{b,gen}$; the periodic oscillations are not harmonic. These oscillations can be presented as the Fourier series $a_{ph}(t) = \sum_n a_n \exp(-i\omega_{b,gen} n t)$, where the coefficients $a_n$ are time-independent. Figure 3(a) shows the dependence of the stationary self-oscillations of the QD polarization $|a_{ph}|$ on time $t$ for eight periods. Note that we use the term "slowly-varying amplitude" for the complex variable $a_{opt} + a_{ph}(t)$, which has the absolute value and the phase. This results in the modulation of the total phase as well as the phase of phonon part $a_{ph}(t)$ (see Fig. 3(b)). The period of the modulation is the same as the period of the modulation of the amplitude of $a_{ph}(t)$. The reason for this matching is that both oscillations are due to auto-oscillations of the optical phonons and the Fröhlich interaction between phonons and active medium polarization.

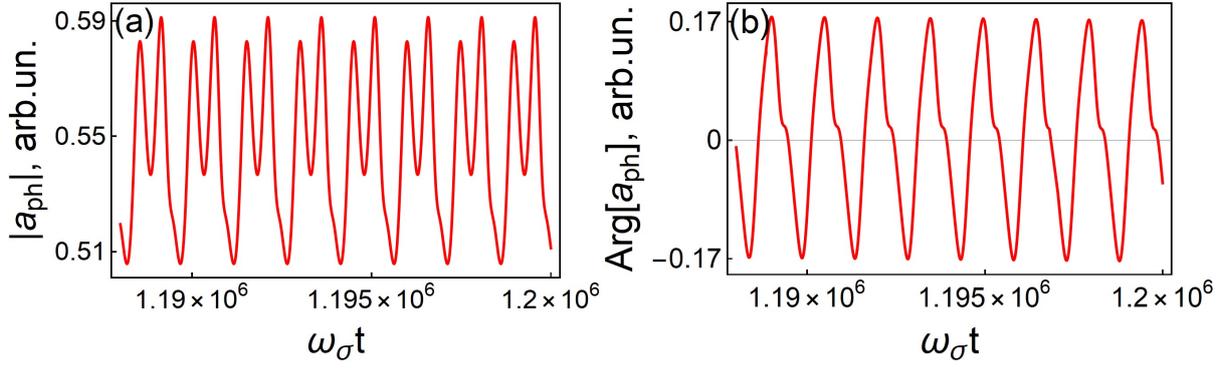

**Figure 3**. Dependencies of (a) the cavity electric field $|a_{ph}(t)|$ and (b) its phase $\arg(a_{ph}(t))$ on time. $D_0 = 0.6$, $g = 0.019 eV$. Other parameters are the same as in Fig. 1.

Figure 4(a) shows the spectrum of the oscillations of $|a_{ph}(t)|$ which is determined by the equality

$$F_{|a_{ph}|}(\omega) = \frac{1}{T}\int_{t_0}^{t_0+T} dt \exp(i\omega t)|a_{ph}(t)|, \quad \omega = n\omega_{b,gen}, \quad n \in \mathbb{Z}. \tag{25}$$

Note that this spectrum determines the spectrum of laser radiation[37,41]. The spectrum contains frequencies that are multiples of the frequency of coherent phonons $\omega_{b,gen}$, which is a consequence of the interdependence between $a$ and $b$. The latter is due to the nonlinearity of the Frohlich interaction between the optical phonon and the exciton.

Figure 4(b) shows the projection of the closed cycle of the self-oscillations on the space $|a|$, $|b|$, and $D$. One can see that the system dynamics is presented by a closed curve which is the manifestation of regime periodicity.

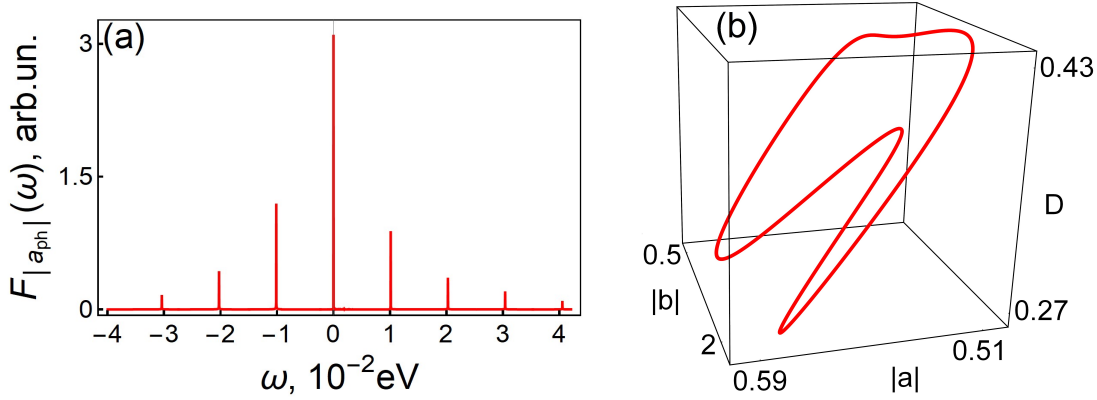

**Figure 4.** (a) The discreet spectrum $F_{|a_{ph}|}(\omega)$, obtained by our computer simulation confirms Eq. (25), $D_0 = 0.6$, $g = 0.019 eV$. (b) The closed cycle of self-oscillation in the space of variables $|a(t)|$, $|b(t)|$, and $|D(t)|$. The parameters are the same as in Fig. 3.

Note that as shown on the plane $(g, D_0)$ (Fig. 1), the laser threshold $D_{th\_opt}$ and the one for self-oscillations with coherent phonons, $D_{th\_ph}(g)$, are different. This means that the excitation of coherent phonons is not a direct consequence of the Raman scattering of the coherent cavity mode.

The appearance of $b_{ph} \neq 0$ implies the emergence of coherent phonons. The latter does not appear in spontaneous Raman scattering even if the incident field is coherent. Thus, this fact cannot be attributed to the Raman scattering of the laser mode. This is confirmed by our computer experiment. Figure 1 shows that the threshold for a coherent harmonic laser mode $D_{th\_opt}$ and the threshold $D_{th\_ph}$ for the appearance of coherent phonons are different.

Usually, the coherent phonons arise in systems at the coherent anti-Stokes Raman spectroscopy (CARS) or after the onset of Raman lasing. In CARS, the appearance of coherent phonons is associated with the resonant excitation of optical phonons at the frequency difference of *two* incident *coherent* waves[18,42,43]. In a conventional laser and in our scheme, the pumping is *incoherent*, and the only coherent wave is the resonator mode above the laser threshold $D_{th\_opt}$. This is clearly not enough for the CARS mechanism to lead to the excitation of coherent phonons.

**The case of detuning $\left|\omega_a - \omega_\sigma^{(g)}\right| > 0$**

As noted above, at $g \neq 0$, instead of $\omega_\sigma$, the frequency of self-oscillations is determined by $\omega_\sigma^{(g)}$, which depends on the values of $g$, $D_{th\_opt}$, and $\omega_a$ (see Eq. (18)). In this subsection, we consider the operational regimes in the case of $\omega_a \neq \omega_\sigma^{(g)}$.

For conventional lasers ($g = 0$), the optimal condition for lasing is matching the resonator frequency $\omega_a$ with the transition frequency $\omega_\sigma$. When this happens, the laser threshold is minimal. With an increase in the difference $\left|\omega_a - \omega_\sigma\right|$, the lasing threshold $D_{th\_opt}$ increases. There is the value of $\left|\omega_a - \omega_\sigma\right|$, at which $D_{th\_opt}$ becomes greater than $D_0$. Then, the pump rate is not sufficient to start lasing, and self-oscillation does not appear. The trivial solution becomes stable again. Consequently, the lasing is absent.

As our computer simulation shows, similar behavior is observed at $g \neq 0$. In this case, however, the role of $\omega_\sigma$ is played by $\omega_\sigma^{(g)}$. At a fixed pump rate $D_0 > D_{th\_opt}$, with an increase in the difference $\left|\omega_a - \omega_\sigma^{(g)}\right|$, the lasing threshold also grows, and sooner or later, this should lead to the suppression of lasing. By computer simulation, we determine for each $g$ the interval of frequencies $\omega_a$, where lasing is observed. The lasing interval for a conventional laser now transforms into the lasing corridor on the plane $(\omega_a, g)$ (see Fig. 5).

The results of the computer simulation are shown in Fig. 5(a). On the plane of the parameters $\omega_a$ and $g$, the dependence $\omega_\sigma^{(g)}(g)$ at fixed $D_0 > D_{th\_opt}$ is shown by the blue solid line. The corridor of parameters for which the lasing is observed (the lasing corridor) is bounded by the dotted lines. Outside of this corridor, only the trivial solution is stable, and the system does not lase. Below we deal with the parameters inside the lasing corridor.

Inside the lasing corridor, a coherent self-oscillation of phonons in the active medium may occur simultaneously with the self-oscillation of the electromagnetic field in the cavity. In Fig. 5(a), the region in which coherent phonons are excited is bounded by the solid black curve. In this figure, the stable solutions with coherent phonons belong to the area colored in orange.

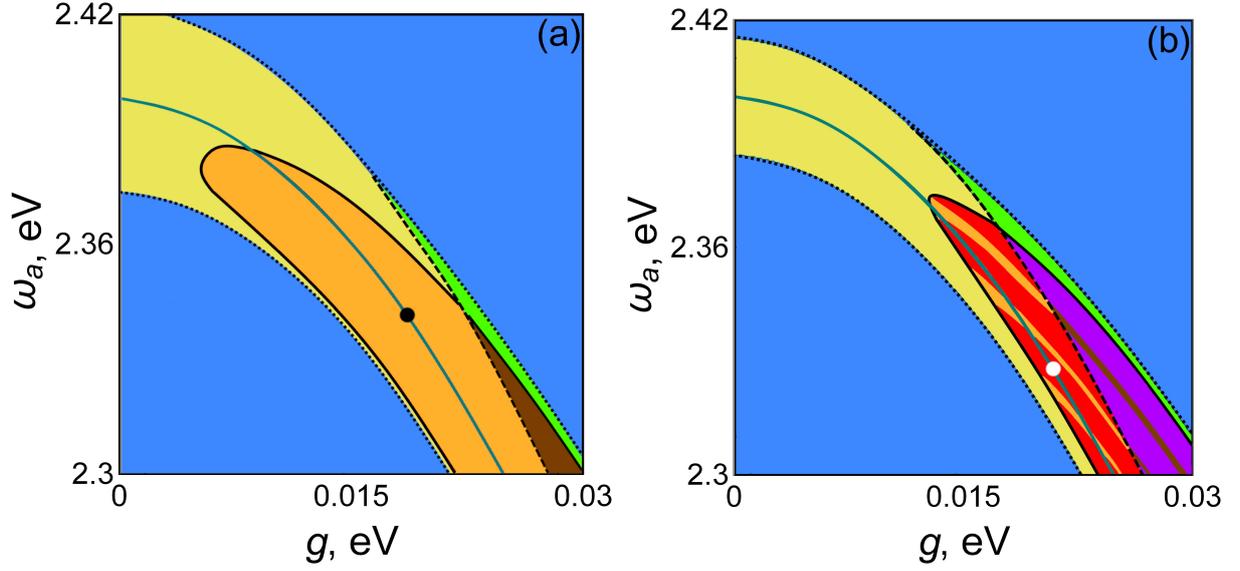

**Figure 5**. Operational regimes of lasing in the parameter plane $(\omega_a, g)$; (a) $\gamma_\sigma = 0.005 eV$ and (b) $\gamma_\sigma = 0.002 eV$. In the the areas colored in blue, lasing is absent; the conventional lasing regime occurs in the yellow area; the operational regime with coherent phonons occupies the orange area, the regime of the chaotic operation occurs in the area colored in red; the area where both trivial and common lasing regimes are stable is colored in green (yellow + blue), the area in which both the common laser regime and the regime with coherent phonons are stable is colored in brown (orange + blue); finally, the area in which both trivial and chaotic solutions are stable is colored in purple (red + blue). The other system parameters are $D_0 = 0.6$, $\Omega_R = 0.01 eV$, $\gamma_b = 0.0002 eV$, $\gamma_a = 0.001 eV$, $\omega_b = 0.01 eV$, $\omega_\sigma = 2.4 eV$, $\gamma_D = 0.001 eV$, and $\gamma_p = 0.004 eV$.

Usually, the system dynamics is uniquely determined by the external parameters $\{\omega_a, g, D_0\}$ and is independent on initial conditions of Eqs. (3) - (6), $\{a, \sigma, D, b\}_{t=0}$. The most intriguing of our results is that there is an area in which two solutions corresponding to different physical regimes are stable. In Fig. 5(a), this area lies between the upper dotted line and the dashed line. Moreover, in the area shown in green, the trivial solution and the periodic harmonic solution are stable, while in the area colored in brown, the trivial and periodic non-harmonic solutions are stable. Here, for different values $\{a, \sigma, D, b\}_{t=0}$, the system arrives at the different stationary state. For example, in the green area, for the initial conditions close to the trivial solution, the system does not begin lasing with time. If the initial conditions are far from the trivial solution, at large times, the conventional laser self-oscillation mode is observed.

It should be noted that each point in Fig. 4 corresponds to a particular system having a specific resonator and active material. The resonator frequency $\omega_a$ and the Fröhlich interaction constant $g$ are plotted along the axes; the Rabi frequency and the pump rate are fixed. Equations (3) – (6) are solved for each such a system.

**The transition from self-oscillations with coherent phonons to chaos**

As noted above, in the absence of the electron-phonon interaction, the laser system could be described by the "gas laser" approximation, $\gamma_a < \gamma_\sigma$, in which the chaotic regime (the Lorenz attractor) does not arise[29]. Thus, there is no reason to expect a transition of the considered system to the strange attractor regime.

Computer simulation shows that the electron-phonon interaction significantly changes the laser dynamics. Namely, if the rate of the exciton dephasing $\gamma_\sigma$ is reduced (note that the inequality $\gamma_a < \gamma_\sigma$ still holds), then in the region where coherent phonons may arise, areas of chaos appear. Such a decrease in $\gamma_\sigma$ can be achieved, for example, by lowering the temperature. For such a case, various operational regimes are shown in Fig. 5(b).

In Fig. 6, time dynamics $|\sigma(t)|$ and its spectrum $F_{|\sigma|}(\omega)$ are shown. The parameters $g$, $D_0$, and $\omega_a$ correspond to the white point in Fig. 5(b). With the accuracy of our computer simulation, which is about $10^{-7}$ eV, all frequencies are present in the spectrum. Such a spectrum drastically differs from the discrete spectrum observed for self-oscillations (Fig. 4a). This is an indication of a chaotic regime.

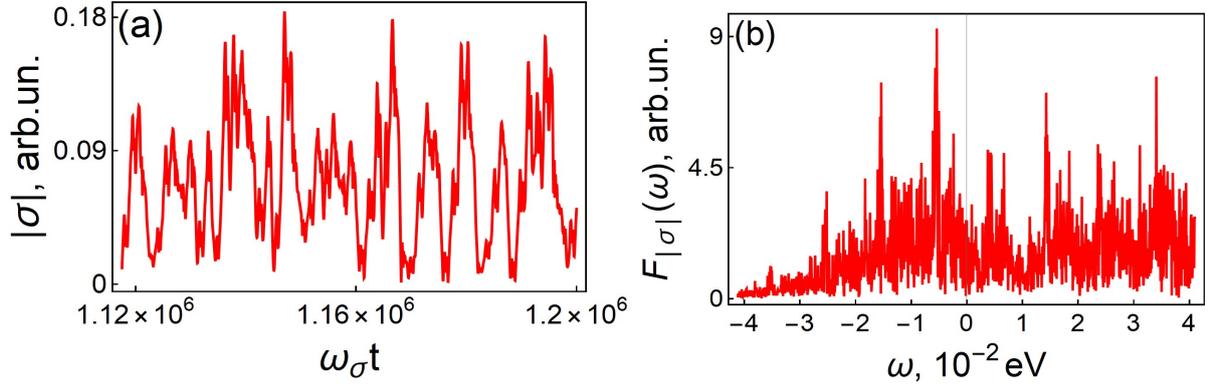

**Figure 6.** Time dynamics $|\sigma(t)|$ (a) and its spectrum $F_{|\sigma|}(\omega)$ corresponding to the white point in Fig. 5(b), in which the chaotic regime is realized. $g = 0.0214 eV$ and $\omega_a = 2.327 eV$; other parameters are the same as in Fig. 5(b).

**Conclusions**

We have considered a conventional single-mode laser consisting of a resonator and a gain medium. Previous studies of conventional laser systems ignored high Raman cross-section of active molecules leading to a strong coupling between excitons and optical phonons. In this work, as far as we know, we consider the effect of the vibrational degrees of freedom on the laser dynamics for the first time.

We describe the exciton-phonon coupling by the Fröhlich Hamiltonian and find that there exists a critical value of the Fröhlich coupling interaction constant $g_{cr}$. Below $g_{cr}$, the conventional laser dynamics is observed, above $g_{cr}$, new operational regimes arise. The value of this constant $g_{cr}$ is about $0.01\,\text{eV}$. This value is realized in several types of quantum dots, such as CdSe, CdS, and PbS. The Raman cross-section for these quantum dots is $\sim 4\cdot 10^{-6}$ Å; it corresponds to the Fröhlich interaction constant $g \simeq 0.025 \text{eV}$, which is of the order of $g_{cr}$.

The parameters $\omega_a$, $g$, and $D_0$ uniquely determine the type of solution to equations of motion (3)-(6). We have done a comprehensive study of the types of solutions for each set of values of the parameters. Different colored regions in Figs. 1 and 5 correspond to the different types of the obtained solutions.

In one of these regimes, a new type of self-oscillations appears. In this regime, both coherent phonons and the coherent optical field are involved. The period of these self-oscillations is equal to $2\pi/\omega_{b,gen}$, where $\omega_{b,gen}$ is the frequency of generated coherent phonons,. The spectrum of these oscillations contains a series of equidistant frequencies that differ by the frequency $\omega_{b,gen}$. The transition to the new self-oscillation regime occurs as the Hopf bifurcation, as the pump rate increases. The new regime is characterized by its own pump rate threshold that differs from the lasing threshold.

Another unusual regime is a chaotic laser dynamics, the spectrum of which contains incommensurable frequencies. This regime is realized for the same values of the Fröhlich interaction constant as the previous regime, but for lesser values of the exciton dephasing rate, $\gamma_\sigma \sim 10^{-3}\,\text{eV}$, that can be achieved, for example, by lowering temperature.

The most unusual property of the new regimes is that above the second threshold, in certain ranges of the resonator frequency, the pump rate, and the Fröhlich constant, more than one solution of the system equations of motion are stable. In this region, the the initial values of the resonator field, the active medium polarization, the population inversion, and the phonon amplitude govern which regime is realized.

In essence, the new self-oscillations described in the paper represent a generator of coherent optical phonons, which may serve as a basis for phonon nanolaser. Such a device can be used for a substantial resolution enhancement in image processing. In addition, the generation of coherent phonons is important for optomechanical applications[27], in quantum information technology[44], and can be used for quantum operations and protocols[45,46].

## Methods

**Master equation for the system density matrix.** To describe the phenomena under consideration, we use the technique of the Lindblad equation. This equation describes the time-evolution of the density matrix[23,24,37]

$$d\hat{\rho}/dt = \frac{i}{\hbar}[\hat{H},\hat{\rho}] + L_{\hat{a}}[\hat{\rho}] + L_{\hat{\sigma}}[\hat{\rho}] + L_{\hat{\sigma}^\dagger}[\hat{\rho}] + L_{\hat{D}}[\hat{\rho}] + L_{\hat{b}}[\hat{\rho}], \qquad (26)$$

where the superoperators

$$L_{\hat{a}}[\hat{\rho}] = \frac{\gamma_a}{2}\left(2\hat{a}\hat{\rho}\hat{a}^\dagger - \hat{a}^\dagger\hat{a}\hat{\rho} - \hat{\rho}\hat{a}^\dagger\hat{a}\right) \tag{27}$$

and

$$L_{\hat{\sigma}}[\hat{\rho}] = \frac{\gamma_D}{2}\left(2\hat{\sigma}\hat{\rho}\hat{\sigma}^\dagger - \hat{\sigma}^\dagger\hat{\sigma}\hat{\rho} - \hat{\rho}\hat{\sigma}^\dagger\hat{\sigma}\right) \tag{28}$$

describe energy relaxations of the cavity mode and the exciton, respectively[24,37]. The superoperator

$$L_{\hat{\sigma}^\dagger}[\hat{\rho}] = \frac{\gamma_p}{2}\left(2\hat{\sigma}^\dagger\hat{\rho}\hat{\sigma} - \hat{\sigma}\hat{\sigma}^\dagger\hat{\rho} - \hat{\rho}\hat{\sigma}\hat{\sigma}^\dagger\right) \tag{29}$$

describes incoherent pumping of the QD exciton[47]. Note that by incoherent pumping we mean the pumping of the high-lying energy levels by a coherent or incoherent light whose frequency is higher than $\omega_\sigma$. Subsequent nonradiative relaxation of electronic states to the exciton state $|e\rangle$ produces the pumping of the exciton.

The superoperator

$$L_{\hat{D}}[\hat{\rho}] = \frac{\gamma_{deph}}{2}\left(2\hat{D}\hat{\rho}\hat{D} - \hat{D}\hat{D}\hat{\rho} - \hat{\rho}\hat{D}\hat{D}\right) \tag{30}$$

describes the relaxation of the phase of the exciton dipole moment. This superoperator is responsible for dephasing – the phase destruction of the exciton dipole moment. The quantities $\gamma_a$, $\gamma_{deph}$, $\gamma_D$, and $\gamma_p$ have the meaning of the rates of respective relaxation and pump processes. The last term in Eq. (26) has the form

$$L_{\hat{b}}[\hat{\rho}] = \frac{\gamma_b}{2}\left(2\hat{b}\hat{\rho}\hat{b}^\dagger - \hat{b}^\dagger\hat{b}\hat{\rho} - \hat{\rho}\hat{b}^\dagger\hat{b}\right). \tag{31}$$

It describes the energy relaxation of optical phonons[24,37].

**Derivation of Eqs. (3) – (6).** We start with Lindblad Eq. (26), which explicit form is

$$\begin{aligned}d\hat{\rho}/dt &= \frac{i}{\hbar}[\hbar\omega_a\hat{a}^\dagger\hat{a} + \hbar\omega_\sigma\hat{\sigma}^\dagger\hat{\sigma} + \hbar\Omega_R\left(\hat{a}\hat{\sigma}^\dagger + \hat{a}^\dagger\hat{\sigma}\right) + \hbar\omega_b\hat{b}^\dagger\hat{b} + \hbar g\hat{\sigma}^\dagger\hat{\sigma}\left(\hat{b}^\dagger + \hat{b}\right), \hat{\rho}] \\ &+ \frac{\gamma_a}{2}\left(2\hat{a}\hat{\rho}\hat{a}^\dagger - \hat{a}^\dagger\hat{a}\hat{\rho} - \hat{\rho}\hat{a}^\dagger\hat{a}\right) + \frac{\gamma_D}{2}\left(2\hat{\sigma}\hat{\rho}\hat{\sigma}^\dagger - \hat{\sigma}^\dagger\hat{\sigma}\hat{\rho} - \hat{\rho}\hat{\sigma}^\dagger\hat{\sigma}\right) \\ &+ \frac{\gamma_p}{2}\left(2\hat{\sigma}^\dagger\hat{\rho}\hat{\sigma} - \hat{\sigma}\hat{\sigma}^\dagger\hat{\rho} - \hat{\rho}\hat{\sigma}\hat{\sigma}^\dagger\right) + \frac{\gamma_{deph}}{2}\left(2\hat{D}\hat{\rho}\hat{D} - \hat{D}\hat{D}\hat{\rho} - \hat{\rho}\hat{D}\hat{D}\right). \\ &+ \frac{\gamma_b}{2}\left(2\hat{b}\hat{\rho}\hat{b}^\dagger - \hat{b}^\dagger\hat{b}\hat{\rho} - \hat{\rho}\hat{b}^\dagger\hat{b}\right).\end{aligned} \tag{32}$$

To obtain equations of motions of operators in Eq. (32), we use the equality

$$\left\langle \frac{d\hat{A}}{dt} \right\rangle \equiv \frac{dA}{dt} = Tr\left(\frac{d\hat{\rho}}{dt}\hat{A}\right)$$

that expresses the expected value of the time-derivative of an operator $\hat{A}$ through the time-derivative of the density matrix. We also use the invariance with respect to cyclic permutation of operators under the trace, the following boson commutation relations $\left[\hat{a},\hat{a}^{\dagger}\right]=1$ and $\left[\hat{b},\hat{b}^{\dagger}\right]=1$ for the resonator mode and optical phonon, and the properties of the two-level system operators: $\hat{\sigma}\hat{D}=\hat{\sigma}$, $\hat{\sigma}^{\dagger}\hat{D}=-\hat{\sigma}^{\dagger}$, $\hat{\sigma}\hat{\sigma}=\hat{\sigma}^{\dagger}\hat{\sigma}^{\dagger}=0$, $\hat{D}\hat{D}=1$, $\hat{\sigma}^{\dagger}\hat{\sigma}=(1+\hat{D})/2$, and $\hat{\sigma}\hat{\sigma}^{\dagger}=(1-\hat{D})/2$. For the expected value of the operator $\hat{a}$ we have

$$\begin{aligned}
da/dt &= Tr(\hat{a}\dot{\hat{\rho}}) = -i\,Tr\left(\hat{a}\omega_{a}\hat{a}^{\dagger}\hat{a}\hat{\rho} + \hat{a}\Omega_{R}\left(\hat{a}\hat{\sigma}^{\dagger}+\hat{a}^{\dagger}\hat{\sigma}\right)\hat{\rho} - \hat{a}\hat{\rho}\omega_{a}\hat{a}^{\dagger}\hat{a} + \hat{a}\hat{\rho}\Omega_{R}\left(\hat{a}\hat{\sigma}^{\dagger}+\hat{a}^{\dagger}\hat{\sigma}\right)\right) \\
&+ \frac{\gamma_{a}}{2}Tr\left(2\hat{a}\hat{a}\hat{\rho}\hat{a}^{\dagger} - \hat{a}\hat{a}^{\dagger}\hat{a}\hat{\rho} - \hat{a}\hat{\rho}\hat{a}^{\dagger}\hat{a}\right) = -i\,Tr\left(\omega_{a}\left[\hat{a},\hat{a}^{\dagger}\hat{a}\right]\hat{\rho} + \Omega_{R}\left[\hat{a},\hat{a}^{\dagger}\hat{\sigma}\right]\hat{\rho}\right) \\
&+ \frac{\gamma_{a}}{2}Tr\left(\left(2\hat{a}^{\dagger}\hat{a}\hat{a} - \hat{a}\hat{a}^{\dagger}\hat{a} - \hat{a}^{\dagger}\hat{a}\hat{a}\right)\hat{\rho}\right) = -i\,Tr\left(\omega_{a}\hat{a}\hat{\rho} + \Omega_{R}\hat{\sigma}\hat{\rho}\right) \\
&+ \frac{\gamma_{a}}{2}Tr\left(\left(\left[\hat{a}^{\dagger},\hat{a}\right]\hat{a}\right)\hat{\rho}\right) = -i\omega_{a}\langle\hat{a}\rangle - i\Omega_{R}\langle\hat{\sigma}\rangle - \frac{\gamma_{a}}{2}\langle\hat{a}\rangle \equiv \left(-i\omega_{a} - \gamma_{a}/2\right)a - i\Omega_{R}\sigma.
\end{aligned} \qquad (33)$$

This is Eq. (3).

For the expected value of the operator $\hat{\sigma}$, we obtain

$$\begin{aligned}
d\sigma/dt &= Tr(\hat{\sigma}\dot{\hat{\rho}}) = -i\,Tr\left(\hat{\sigma}\omega_{\sigma}\hat{\sigma}^{\dagger}\hat{\sigma}\hat{\rho} + \hat{\sigma}\Omega_{R}\left(\hat{a}\hat{\sigma}^{\dagger}+\hat{a}^{\dagger}\hat{\sigma}\right)\hat{\rho} + \hat{\sigma}g\hat{\sigma}^{\dagger}\hat{\sigma}\left(\hat{b}^{\dagger}+\hat{b}\right)\hat{\rho}\right) \\
&+ i\,Tr\left(\hat{\sigma}\hat{\rho}\omega_{\sigma}\hat{\sigma}^{\dagger}\hat{\sigma} + \hat{\sigma}\hat{\rho}\Omega_{R}\left(\hat{a}\hat{\sigma}^{\dagger}+\hat{a}^{\dagger}\hat{\sigma}\right) + \hat{\sigma}\hat{\rho}g\hat{\sigma}^{\dagger}\hat{\sigma}\left(\hat{b}^{\dagger}+\hat{b}\right)\right) \\
&+ \frac{\gamma_{D}}{2}Tr\left(2\hat{\sigma}\hat{\sigma}\hat{\rho}\hat{\sigma}^{\dagger} - \hat{\sigma}\hat{\sigma}^{\dagger}\hat{\sigma}\hat{\rho} - \hat{\sigma}\hat{\rho}\hat{\sigma}^{\dagger}\hat{\sigma}\right) + \frac{\gamma_{p}}{2}Tr\left(2\hat{\sigma}\hat{\sigma}^{\dagger}\hat{\rho}\hat{\sigma} - \hat{\sigma}\hat{\sigma}\hat{\sigma}^{\dagger}\hat{\rho} - \hat{\sigma}\hat{\rho}\hat{\sigma}\hat{\sigma}^{\dagger}\right) \\
&+ \frac{\gamma_{deph}}{4}Tr\left(2\hat{\sigma}\hat{D}\hat{\rho}\hat{D} - \hat{\sigma}\hat{D}\hat{D}\hat{\rho} - \hat{\sigma}\hat{\rho}\hat{D}\hat{D}\right) \\
&= -i\,Tr\left(\omega_{\sigma}\hat{\sigma}\hat{\rho} + \hat{\sigma}\Omega_{R}\hat{a}\left[\hat{\sigma},\hat{\sigma}^{\dagger}\right]\hat{\rho} + g\hat{\sigma}\left(\hat{b}^{\dagger}+\hat{b}\right)\hat{\rho}\right) + i\,Tr\left(\Omega_{R}\hat{a}\hat{\sigma}^{\dagger}\hat{\sigma}\hat{\rho}\right) \\
&- \frac{\gamma_{D}}{2}Tr\left(\hat{\sigma}\hat{\sigma}^{\dagger}\hat{\sigma}\hat{\rho}\right) - \frac{\gamma_{p}}{2}Tr\left(\hat{\sigma}\hat{\sigma}^{\dagger}\hat{\sigma}\hat{\rho}\right) - \frac{\gamma_{deph}}{4}Tr\left(4\hat{\sigma}\hat{\rho}\right) \\
&= -i\omega_{\sigma}Tr(\hat{\sigma}\hat{\rho}) - ig\,Tr\left(\left(\hat{b}+\hat{b}^{\dagger}\right)\hat{\sigma}\hat{\rho}\right) + i\Omega_{R}Tr\left(\hat{a}\hat{D}\hat{\rho}\right) - \left(\frac{\gamma_{D}}{2}+\frac{\gamma_{p}}{2}+\gamma_{deph}\right)Tr(\hat{\sigma}\hat{\rho}) \\
&= \left(-i\omega_{\sigma} - \gamma_{\sigma}\right)\sigma - ig\left\langle\left(\hat{b}+\hat{b}^{\dagger}\right)\hat{\sigma}\right\rangle + i\Omega_{R}\langle\hat{a}\hat{D}\rangle.
\end{aligned} \qquad (34)$$

At the last step in obtaining Eq. (34), we introduce the notation $\gamma_\sigma = \gamma_D/2 + \gamma_p/2 + \gamma_{deph}$. Further, we neglect the correlations between pairs of operators $\hat{b} + \hat{b}^\dagger$ and $\hat{\sigma}$, $\hat{a}$ and $\hat{D}$, assuming that $\langle (\hat{b} + \hat{b}^\dagger)\hat{\sigma} \rangle = \langle \hat{b} + \hat{b}^\dagger \rangle \langle \hat{\sigma} \rangle = (b + b^*)\sigma$. As a result, we obtain

$$d\sigma/dt = \operatorname{Tr}(\hat{\sigma}\dot{\hat{\rho}}) = \left(-i\left(\omega_\sigma + g(b + b^*)\right) - \gamma_\sigma\right)\sigma + i\Omega_R aD. \tag{35}$$

This is Eq. (4)

For the expected value of the operator $\hat{D}$, we have

$$\begin{aligned} dD/dt &= \operatorname{Tr}(\hat{D}\dot{\hat{\rho}}) = -i\operatorname{Tr}\left(\hat{D}\omega_\sigma \hat{\sigma}^\dagger \hat{\sigma}\hat{\rho} + \hat{D}\Omega_R(\hat{a}\hat{\sigma}^\dagger + \hat{a}^\dagger\hat{\sigma})\hat{\rho} + \hat{D}g\hat{\sigma}^\dagger\hat{\sigma}(\hat{b}^\dagger + \hat{b})\hat{\rho}\right) \\ &+ i\operatorname{Tr}\left(\hat{D}\hat{\rho}\omega_\sigma\hat{\sigma}^\dagger\hat{\sigma} + \hat{D}\hat{\rho}\Omega_R(\hat{a}\hat{\sigma}^\dagger + \hat{a}^\dagger\hat{\sigma}) + \hat{D}\hat{\rho}g\hat{\sigma}^\dagger\hat{\sigma}(\hat{b}^\dagger + \hat{b})\right) \\ &+ \frac{\gamma_D}{2}\operatorname{Tr}\left(2\hat{D}\hat{\sigma}\hat{\rho}\hat{\sigma}^\dagger - \hat{D}\hat{\sigma}^\dagger\hat{\sigma}\hat{\rho} - \hat{D}\hat{\rho}\hat{\sigma}^\dagger\hat{\sigma}\right) + \frac{\gamma_p}{2}\operatorname{Tr}\left(2\hat{D}\hat{\sigma}^\dagger\hat{\rho}\hat{\sigma} - \hat{D}\hat{\sigma}\hat{\sigma}^\dagger\hat{\rho} - \hat{D}\hat{\rho}\hat{\sigma}\hat{\sigma}^\dagger\right) \\ &+ \frac{\gamma_{deph}}{4}\operatorname{Tr}\left(2\hat{D}\hat{D}\hat{\rho}\hat{D} - \hat{D}\hat{D}\hat{D}\hat{\rho} - \hat{D}\hat{\rho}\hat{D}\hat{D}\right) \\ &= -i\operatorname{Tr}\left(\Omega_R \hat{a}\hat{\sigma}^\dagger\hat{\rho} - \Omega_R \hat{a}^\dagger\hat{\sigma}\hat{\rho}\right) + i\operatorname{Tr}\left(-\Omega_R \hat{a}\hat{\sigma}^\dagger\hat{\rho} + \Omega_R \hat{a}^\dagger\hat{\sigma}\hat{\rho}\right) \\ &+ \frac{\gamma_D}{2}\operatorname{Tr}\left(2\hat{\sigma}^\dagger\hat{D}\hat{\sigma}\hat{\rho} - \hat{D}\hat{\sigma}^\dagger\hat{\sigma}\hat{\rho} - \hat{\sigma}^\dagger\hat{\sigma}\hat{D}\hat{\rho}\right) + \frac{\gamma_p}{2}\operatorname{Tr}\left(2\hat{\sigma}\hat{D}\hat{\sigma}^\dagger\hat{\rho} - \hat{D}\hat{\sigma}\hat{\sigma}^\dagger\hat{\rho} - \hat{\rho}\hat{\sigma}\hat{\sigma}^\dagger\hat{D}\right) \\ &= 2i\Omega_R\operatorname{Tr}\left((\hat{a}^\dagger\hat{\sigma} - \hat{a}\hat{\sigma}^\dagger)\hat{\rho}\right) - \frac{\gamma_D}{2}\operatorname{Tr}(4\hat{\sigma}^\dagger\hat{\sigma}\hat{\rho}) + \frac{\gamma_p}{2}\operatorname{Tr}(4\hat{\sigma}\hat{\sigma}^\dagger\hat{\rho}) \\ &= 2i\Omega_R\langle \hat{a}^\dagger\hat{\sigma} - \hat{a}\hat{\sigma}^\dagger \rangle - \gamma_D\langle 1 + \hat{D} \rangle + \gamma_p\langle 1 - \hat{D} \rangle = -(\gamma_D + \gamma_p)(D - D_0) + 2i\Omega_R\langle \hat{a}^\dagger\hat{\sigma} - \hat{a}\hat{\sigma}^\dagger \rangle. \end{aligned} \tag{36}$$

In this equation, we introduce the notation $D_0 = (\gamma_p - \gamma_D)/(\gamma_p + \gamma_D)$. In deriving Eq. (36), we suppose that $\langle \hat{a}\hat{\sigma}^\dagger \rangle = \langle \hat{a} \rangle\langle \hat{\sigma}^\dagger \rangle = a\sigma^*$ and $\langle \hat{a}^\dagger\hat{\sigma} \rangle = \langle \hat{a}^\dagger \rangle\langle \hat{\sigma} \rangle = a^*\sigma$. As a result, we obtain

$$dD/dt = \operatorname{Tr}(\hat{D}\dot{\hat{\rho}}) = -(\gamma_D + \gamma_p)(D - D_0) + 2i\Omega_R(a^*\sigma - a\sigma^*). \tag{37}$$

This is Eq. (6).

Finally, for the expected value of the operator $\hat{b}$, we have

$$db/dt = \text{Tr}(\hat{b}\dot{\hat{\rho}}) = -i\,\text{Tr}\left(\hat{b}\omega_b \hat{b}^\dagger \hat{b}\hat{\rho} + \hat{b}g\hat{\sigma}^\dagger \hat{\sigma}(\hat{b}+\hat{b}^\dagger)\hat{\rho} - \hat{b}\hat{\rho}\omega_b \hat{b}^\dagger \hat{b} - \hat{b}\hat{\rho}g\hat{\sigma}^\dagger \hat{\sigma}(\hat{b}+\hat{b}^\dagger)\right)$$

$$+\frac{\gamma_b}{2}\text{Tr}\left(2\hat{b}\hat{b}\hat{\rho}\hat{b}^\dagger - \hat{b}\hat{b}^\dagger \hat{b}\hat{\rho} - \hat{b}\hat{\rho}\hat{b}^\dagger \hat{b}\right) = -i\,\text{Tr}\left(\omega_b\left[\hat{b},\hat{b}^\dagger \hat{b}\right]\hat{\rho} + g\hat{\sigma}^\dagger \hat{\sigma}\left[\hat{b},\hat{b}^\dagger\right]\hat{\rho}\right)$$

$$+\frac{\gamma_b}{2}\text{Tr}\left(\left(2\hat{b}^\dagger \hat{b}\hat{b} - \hat{b}\hat{b}^\dagger \hat{b} - \hat{b}^\dagger \hat{b}\hat{b}\right)\hat{\rho}\right) = -i\,\text{Tr}\left(\omega_b \hat{b}\hat{\rho} + g\hat{\sigma}^\dagger \hat{\sigma}\hat{\rho}\right) \quad (38)$$

$$+\frac{\gamma_b}{2}\text{Tr}\left(\left(\left[\hat{b}^\dagger,\hat{b}\right]\hat{b}\right)\hat{\rho}\right) = -i\omega_b\langle\hat{b}\rangle - ig\langle 1+\hat{D}\rangle/2 - \frac{\gamma_b}{2}\langle\hat{b}\rangle \equiv (-i\omega_b - \gamma_b/2)b - ig(1+D)/2.$$

This is Eq. (5).

**Maxwell-Bloch equations in the absence of interaction between exciton and optical phonon.**
We are interested in the expected value of the field annihilation operator in the resonator $\langle\hat{a}\rangle = a$, which is proportional to the average field in the resonator, the expected values of the total exciton dipole moment operator, $\sigma = \langle\hat{\sigma}\rangle$, and the population inversion $\langle\hat{D}\rangle = D$ of the excited level. Using Lindblad equation (26) we obtain the equations of motion for the expected values of the operators $\hat{a}$, $\hat{a}^\dagger$, $\hat{\sigma}$, and $\hat{\sigma}^\dagger$ [23,33]:

$$da/dt = (-i\omega_a - \gamma_a/2)a - i\Omega_R \sigma, \quad (39)$$

$$d\sigma/dt = (-i\omega_\sigma - \gamma_\sigma/2)\sigma + i\Omega_R aD, \quad (40)$$

$$dD/dt = -(\gamma_p + \gamma_D)(D-D_0) + 2i\Omega_R(a^\dagger \sigma - \sigma^\dagger a), \quad (41)$$

where $D_0 = (\gamma_p - \gamma_D)/(\gamma_p + \gamma_D)$.

The trivial stationary solution to system of Eqs. (39) – (41) is: $a = \sigma = 0$, $D = D_0$ [48]. This solution is stable as long as $D_0 < D_{0,th}$ [23], where

$$D_{0,th} = \frac{\gamma_a \gamma_\sigma}{4\Omega_R^2}\left(1 + \frac{4(\omega_a - \omega_\sigma)^2}{(\gamma_a + \gamma_\sigma)^2}\right). \quad (42)$$

For $D \geq D_{th}$, this solution becomes unstable, and a new stable solution at times $t \gg \gamma_a^{-1}, \gamma_\sigma^{-1}, \gamma_v^{-1}$ arises (the Hopf bifurcation)

$$a(t) = \sqrt{(\gamma_p + \gamma_D)(D_0 - D_{0,th})/2\gamma_a}\,e^{-i\omega_{gen}t},$$

$$\sigma(t) = \frac{i(\omega_{gen} - \omega_a) - \gamma_a/2}{\Omega_R}\sqrt{\frac{(\gamma_p + \gamma_D)}{2\gamma_a}(D_0 - D_{0,th})}\,e^{-i\omega_{gen}t}, \quad (43)$$

$$D = D_{th\_opt},$$

where the self-oscillation frequency $\omega_{gen}$ is determined by the "frequency pulling" formula[23,24,40]:

$$\omega_{gen} = \frac{\omega_\sigma \gamma_\sigma + \omega_a \gamma_a}{\gamma_\sigma + \gamma_a}. \qquad (44)$$

In the general case, this self-oscillating solution becomes unstable if two conditions, $\gamma_a > \gamma_D + \gamma_\sigma$ and $D_0 > D_{th}\gamma_a(\gamma_a + \gamma_D + \gamma_\sigma)/\gamma_\sigma(\gamma_a - \gamma_D - \gamma_\sigma)$, are met. In this case, a regime of deterministic chaos arises, which in the phase space of variables, $\{a, \sigma, D\}$ takes the form of the Lorentz attractor [23,49]. But, since we consider the case of a "gas laser" ($\gamma_a \ll \gamma_\sigma$)[29], even the first of the conditions listed above is never satisfied, and there is no reason to expect a transition to the strange attractor regime.

**Effective change in the exciton frequency due to a constant displacement of the QD nuclei under incoherent pumping.** We begin with considering the QD dynamics in the absence of an external field, including the resonator field. This helps us to interpret the regimes of the QD generation in the cavity more clearly.

In the absence of an external field and a resonator, we can put $\Omega_R = 0$. Equations (3) - (6) are reduced to

$$da/dt = (-i\omega_a - \gamma_a/2)a, \qquad (45)$$

$$d\sigma/dt = (-i\omega_\sigma - \gamma_\sigma/2)\sigma - ig(b + b^\dagger)\sigma, \qquad (46)$$

$$db/dt = (-i\omega_b - \gamma_b/2)b - ig(D+1)/2, \qquad (47)$$

$$dD/dt = -(\gamma_p + \gamma_D)(D - D_0). \qquad (48)$$

The trivial stationary solution to this system has the form:

$$a_{st} = 0, \ \sigma_{st} = 0, \ D_{st} = D_0, \ b_{st} = ig\frac{D_0 + 1}{2(-i\omega_b - \gamma_b/2)}. \qquad (49)$$

Let us consider the dynamics of small deviations of the QD dipole moment $\sigma$ from this equilibrium position. Substituting $\sigma = \sigma_{st} + \delta\sigma$, $b = b_{st} + \delta b$, and $D = D_{st} + \delta D$ into Eqs. (45) – (48), assuming $\omega_\sigma \gg g$ (that is always satisfied, since $\omega_\sigma$ lies in the optical range and $g$ is in the terahertz range), and implying that the value of $\delta b$ is small in comparison with $b_{st}$, we obtain

$$d(\delta\sigma)/dt = \left(-i\omega_\sigma - \gamma_\sigma/2\right)\delta\sigma - ig\left(b_{st} + b_{st}^\dagger\right)\delta\sigma = \left(-i(\omega_\sigma + \omega_{sh}) - \gamma_\sigma/2\right)\delta\sigma,, \qquad (50)$$

where we have introduced the notation for the phonon effective frequency shift of the QD[50]:

$$\omega_{sh}^{(trivial)} = -\frac{\omega_b g^2 (D_0 + 1)}{\omega_b^2 + \gamma_b^2/4}. \qquad (51)$$

Thus, the oscillations of the exciton dipole moment near the equilibrium position occur not at the exciton eigenfrequency $\omega_\sigma$ but at the shifted frequency $\omega_\sigma + \omega_{sh}$. The displacement $\omega_{sh}$ is always negative and proportional to the value $D_0 + 1$, which has the meaning of the population of the excited state of the exciton at a nonzero pump rate, $D_0 > -1$.

## Acknowledgments

E.S.A. and A.A.Z. thank a grant from Russian Science Foundation (project No. 20-72-10057) for financial support. E.S.A. thanks foundation for the advancement of theoretical physics and mathematics "Basis." A.A.L. acknowledges the support of the ONR under Grant No. N00014-20-1-2198.

## Author contributions

E.A.T., A.A.Z., E.S.A., and A.A.L made theoretical analysis. E.A.T. performed numerical calculations. E.A.T., E.S.A., A.P.V., and A.A.L. analyzed the results and wrote the manuscript with input from all the authors. All authors contributed to the idea of the work. All authors reviewed the manuscript.

## Competing interests

The authors declare no competing interests.

## Additional information

**Correspondence** and requests for materials should be addressed to E.S.A.